\documentstyle[12pt]{article}
\begin{document}
\newcommand {\be}{\begin{equation}}
 \newcommand {\ee}{\end{equation}}
 \newcommand {\bea}{\begin{array}}
 \newcommand {\cl}{\centerline}
 \newcommand {\eea}{\end{array}}
 \renewcommand {\theequation}{\thesection.\arabic{equation}}
 \newcommand {\newsection}{\setcounter{equation}{0}\section}
 \def\nct{noncommutative torus }
 \def\EOM{equation of motion}
 \def\ncy{noncommutativity }
 \def\nc{noncommutative }
 \def\quan{quantization }
 \def\BC{boundary condition }
 \def\BCs{boundary conditions }
 \def\cons{constraints }
  \baselineskip 0.6 cm
\begin{titlepage}
\vspace{-10mm}

\begin{center}
\begin{large}
{\bf The Reduced Phase Space  \\ of An Open String \\in  The Background B-Field  }\\
\end{large}
\vspace{5mm}
  {\bf M. Dehghani \footnote{e-mail:
mdehghani@ph.iut.ac.ir }} , {\bf A. Shirzad \footnote{e-mail:
shirzad@ipm.ir}}

{\it Department of  Physics, Isfahan University of Technology \\
Isfahan,  IRAN, \\
Institute for Studies in Theoretical Physics and Mathematics\\
P. O. Box: 5531, Tehran, 19395, IRAN.} \vspace{0.3cm}
\end{center}
\abstract{ The problem of an open string in background $B$-field
is discussed. Using the discretized model in details we show that
the system is influenced by infinite number of second class
constraints. We interpret the allowed Fourier modes as the
coordinates of the reduced   phase space. This enables us to
compute the Dirac brackets more easily. We prove that the
coordinates of the string are non-commutative at the boundaries.
We argue that in order to find the Dirac bracket or commutator
algebra of the physical variables, one should not expand the
fields in terms of the solutions of the equations of motion.
Instead, one should impose the set of constraints in suitable
coordinates.

 \vfill}

\end{titlepage}
\newsection{Introduction \label{section1}}
The constrained systems were first introduced by Dirac
\cite{dirac} within the discussion about singular Lagrangians. In
such systems, Euler-Lagrange equations of motion  lead to some
acceleration-free identities as consequences of the singularity of
Hessian (second derivative of Lagrangian with respect to the
velocities). At the Hamiltonian level, this leads to emergence of
constraints, i.e. functions of phase space coordinates which
should vanish on-shell.

 We call that the first class constraints are responsible for gauge transformations;
while second class constraints restrict the system to a smaller
sub-manifold of the phase space in which a Poisson structure  can
be recognized. In a very simplified picture, the first class
constraints may be visualized as some momenta, so that they are
involutive and generate transformations in their conjugate
coordinates (i.e. the gauged variables). On the other hand, the
second class constraints may be visualized as conjugate pairs of
coordinates and momenta whose Poisson brackets in the original
phase space  are nonzero. This seems to contradict the fact that
they vanish on the surface of motion.  Hence, it is necessary to
introduce a new bracket, i.e. the Dirac bracket,  such that the
physical quantities (quantities defined on the constraint surface)
have vanishing brackets with the constraints. A complete review on
constrained systems can be found in \cite{HENOUX}

The singularity of the Lagrangian is not necessarily the only
origin of the constraints. One may find  constraints in the
Lagrangian formalism in any way that one can impose
acceleration-free equations on the system; or in the  Hamiltonian
formalism, in any way that one can impose a primary constraint on
the system. As far as we know, apart from artificial problems in
which primary Hamiltonian constraints are  imposed by hand, no
serious model has been introduced in which constraints emerge
naturally, \textit{but not as the result of the singularity of the
system}.

In the latest years of the last millennium, in another branch of
research, a new phenomenon was discovered which has relationships
with the constrained systems. It was found that for an open string
coupled to a background $B$-field,  the canonical quantization
procedure fails at the end points \cite{AAS2,AAS3,CH2}. It was
also observed in Refs. \cite{SW,VS} based on the work of Ref.
\cite{NAPPI}, that in the presence of mixed boundary conditions
due to $B$-field, the propagators of coordinate fields possess
singularities which can be interpreted as the non-commutativity of
coordinate fields specially at the boundaries (branes). The origin
of this non-commutativity is intensively discussed from then on
\cite{SS,KIM,BRAGA,TL1,TL2,koka,baner1,ZHAO}.

 This observation led some authors  \cite{AAS3,CH2,SS} to
 deduce that the Dirac mechanism of second class
constraints has some role at the boundaries of the string, hence
the idea of considering the boundary conditions as Dirac
constraints was born \cite{AAS3,CH2}. Since  the boundary
conditions just put limitations on the solutions of equations of
motion;  if accepted  as constraints, they should somehow be
related to the second class systems.

In the next section we will review the main set up of the above
model, following mainly  the method of reference \cite{SS}.
Assuming the boundary conditions as constraints, we use the total
Hamiltonian to impose their consistency. As we will see, the
Lagrange multipliers are determined  at the first stage of
consistency (to be zero), but at the same time the second level
constraints emerge. Then third level constraints are derived by
demanding the consistency of second level ones, and so forth. In
this way the procedure of finding the constraints proceeds
unlimitedly. This suggests a new category in comparison with the
ordinary theory of constrained systems in which the constraint
chains terminate after a limited number of steps either by
arriving  at an identity (for first class systems), or by
determining the Lagrange multipliers (for second  class systems)
\cite{LS}.

During recent years several different and even opposite approaches
have been applied to the problem, though most of them have derived
similar results. Some authors do not consider the consistency of
the constraints completely. They calculate the Dirac brackets just
by using primary or at most second level constraints
\cite{KIM,zheng,L}. The idea of infinite number of second class
constraints at the boundaries, however, was accepted by many
authors \cite{AAS3,CH2,SS}, although fewer people note what
happens to the Lagrange multipliers. Some authors have also tried
to overcome the problem by considering discretized model \cite{SS}
or by following the symplectic approach
\cite{BRAGA,TEZUKA,J.JING}. Such approaches, though verify that
the system is constrained in some sense, do not give final
justification about the constraint characteristics of the system.

Our first objective in this work is to find a distinctive
understanding about the constraint structure of the model. We want
to know  in  what sense we find infinite number of constraints. We
think that the key to study this part of the problem is the
discretized model. It is known that the primary constraints emerge
as the continuum limit of the equations of motion of the end
points in the discretized  model \cite{SS,KIM}. However
investigating the consistency conditions to deduce the set of
secondary constraints has not been studied in discretized model
yet. In sections (\ref{section3}) and (\ref{section4}) we have
done this. We will see that considering the continuum limit
together with the physical condition of continuity gives the set
of desired constraints. We think that in the light of  this study
of discrete model  any doubt about the existence of infinite
number of constraints disappears.

Next problem concerns the properties of the reduced phase space.
Specially, one needs to know about the induced brackets on this
space, which is the same as the Dirac brackets. With infinite
number of constraints, the matrix of Poisson brackets of
constraints is infinite dimensional. One needs to invert it to
find the Dirac brackets. Some authors, which have accepted the
existence of infinite number of constraints  \cite{CH2,SS}, have
tried to solve this problem directly. However, as we will explain
in the following section, the mathematical manipulations given in
different papers are not convincing yet. In fact, since people
excepted to end up with non-commutativity in brane coordinates
upon physical intuitions, most of the authors (except a few of
them)  have derived the famous result of Ref. \cite{SW} with more
or less problematic mathematical methods.

In section (\ref{section5}) we will try to give a very simple
approach to find the Dirac brackets on the basis of Fourier modes.
We show that this powerful physical tool, (considered in some
different sense  in Ref. \cite{CH1}) serves as the  suitable
coordinates for describing the reduced phase space. From this
point of view, after imposing the constraints, the brackets of the
remaining physical modes  can be written automatically; and then
the Dirac brackets of the original fields can be derived, using
their expansion in terms of physical modes. Apart from
convenience in calculations, this method helps one understand
better the mechanism which leads to unusual brackets of
coordinate fields at the boundaries.

In section(\ref{section6}) we give a comprehensive discussion
about using the classical equations of motion of the fields in
the process of quantization. We argue that in fact, the algebra
of the observables is the essential entity in the quantization
process rather than their dynamics. We show that just the
dynamics of constraints is essential to be investigated before
quantization. In this sense, we have derived the quantum
properties of the reduced phase space, including the
non-commutativity of brane coordinates, without using the
solutions of the equations of motion.

\newsection{Problem setup \label{section2}}
Consider an open string with coordinate fields $X^{\mu}$  living
in a target space specified by $\mu=0,1,...,D$. Suppose that the
string is coupled to a given antisymmetric field $\cal{F}_{\mu
\nu}$ which for simplicity we assume no dynamics for it. The end
points $\sigma=0, l$ of the string are constrained to move on a
$p+1$ dimensional $D_p$-brane characterized by $X^a=0$ for
$a=p+1,...,D$. The $U(1)$ gauge fields, $A_i$ $i=0,1,...,p$ are
also coupled to the string on the boundary ($D_p$-brane). The
action of the model can now be written as
  \be S=\frac{1}{4\pi
\alpha'}\int_{\Sigma} d\sigma d\tau \left[G_{\mu \nu}\partial_a
X^{\mu}\partial_b X^{\nu}g^{ab}+\epsilon^{ab}B_{\mu \nu}\partial_a
X^{\mu}\partial_b X^{\nu}\right]+\frac{1}{2\pi
\alpha'}\oint_{\partial\Sigma}d\tau A_i\partial_{\tau}X^i\ee where
$\partial\Sigma$ is the boundary of target  space $\Sigma$,
$g_{ab}$ is the metric of world-sheet and $\epsilon^{ab}$ is the
antisymmetric tensor on the world-sheet. We considered for
simplicity only the bosonic sector. Similar arguments can be
applied to a superstring \cite{TL2,CH3}. The  bulk and boundary
fields can be combined to construct modified Born-Infeld field
strength $ {\cal{F}}= B-d A $. Let also both end points attach to
the same brane. Assume $\alpha'=\frac{1}{2 \pi}$ and the
background metric to be flat: $G_{\mu \nu}=\eta_{\mu \nu}$.
Suppose, moreover, that the field strength is constant everywhere,
given by the constant antisymmetric matrix $B_{ij}$. Using  these
simplifications the action reads: \be S=\frac{1}{2}\int d\sigma
d\tau \left[\partial_a X^{\mu}\partial_b X_{\mu}g^{ab}
+B_{ij}\partial_a X^{i}\partial_b X^{j}\epsilon^{ab} \right] \ee
 fixing the diffeomorphisms and scaling invariance, the above
 action takes the form
  \be S=\frac{1}{2}\int d\sigma d\tau
  \left[\dot{X}^{\mu}\dot{X}_{\mu}-X'^{\mu}X'_{\mu}+2B_{ij}\dot{X}^iX'^j
  \right] \label{original}\ee
 where ``dot" and ``prime" represent differentiating with respect to
$\tau$ and $\sigma$ respectively. Besides the conformal symmetry,
this action has a global symmetry under transforming the
coordinate fields by a constant, i.e.
 \be X^{\mu}(\sigma,\tau)\rightarrow X^{\mu}(\sigma,\tau)+c^{\mu}.
 \label{global}\ee
Varying the action with respect to the fields $X^{\mu}$ gives the
equations of motion
  \be \partial^2_{\tau}X^{\mu}(\sigma,\tau)-
  \partial^2_{\sigma} X^{\mu}(\sigma,\tau)=0 \;\;\
  \mu=0,1,...,D \label{EOM}\ee
  together with the boundary conditions
  \be \begin{array}{ll}
\partial_{\sigma} X^i(\sigma,\tau)+B_{ij}\partial_{\tau}
X^j(\sigma,\tau)=0 \hspace{7mm}& i=0,1,\cdots,p \\
X^a=0 & a=p+1,\cdots,D\\
\end{array}
   \label{pri}\ee
   at the end points $\sigma=0, l$. As is apparent the boundary
conditions in the directions perpendicular to the $D_p$-brane are
simply of Dirichlet  type; while along the $D_p$-brane we have
mixed boundary conditions. This is the only place where the
effect of the $B$-field is experienced. Let concentrate in the
following on this part of the problem and ignore the coordinate
fields with Dirichlet boundary conditions. One simple way to
realize this point is to assume that $p=D$. On the other hand, at
the end of calculations one can turn off the $B$-field in as many
direction as one desires to achieve  results concerning the
Neumann boundary conditions. The equations (\ref{pri}) do not
contain accelerations, so it may be viewed as  primary Lagrangian
constraints. The canonical momentum fields and the Hamiltonian of
the system read
  \be \Pi_i(\sigma,\tau)=\partial_{\tau} X(\sigma,\tau)+B_{ij}\partial_{\sigma}X_j(\sigma,\tau)
  \label{momentun}\ee
  \be H=\frac{1}{2}\int_0^l\left[(\Pi_i-B_{ij}\partial_{\sigma}X_j)^2
  +(\partial_{\sigma}X_i)^2\right]d \sigma. \label{hamiltoni}\ee
  The primary constraints (\ref{pri}) in terms of the phase space
  variables are
  \be \Phi_i=\Phi_i(\sigma)|_{\sigma=0}, \;\;\ \bar{\Phi}_i=\Phi_i(\sigma)|_{\sigma=l} \label{primary} \ee
  with
  \be \Phi_i(\sigma,\tau)=M_{ij}\partial_{\sigma}X_j(\sigma,\tau)+B_{ij}\Pi_j(\sigma,\tau),
  \;\;\ M=1-B^2. \label{2def}\ee
 As for every constrained system, given the primary constraints
 (\ref{primary}) the total Hamiltonian
 \be H_T=H+\lambda_i\Phi_i^1+\bar{\lambda}_i\bar{\Phi}_i^1
 \label{hamiltot} \ee
 is responsible for the dynamics of the system. Hence, the
 consistency of the constraints gives

 \be
\begin{array}{l}
  \dot{\Phi}_i^1=\{\Phi_i^1,H_T\}=\{\Phi_i^1,H
 \}+\lambda_j\{\Phi_i^1,\Phi_j^1\}=0  \\
  \dot{\bar{\Phi}}_i^1=\{\bar{\Phi}_i^1,H_T\}=\{\bar{\Phi}_i^1,H
 \}+\bar{\lambda}_j\{\bar{\Phi}_i^1,\bar{\Phi}_j^1\}=0.  \\
\end{array}
 \label{consistant2}\ee
 Using Eqs. (\ref{hamiltoni}-\ref{2def}) and the fundamental Poisson brackets
 \be
\begin{array}{l}
  \{X_i(\sigma,\tau),X_j(\sigma ',\tau)\}=0 \\
  \{X_i(\sigma,\tau),\Pi_j(\sigma ',\tau)\}=\delta_{ij}\delta(\sigma-\sigma ') \\
  \{\Pi_i(\sigma,\tau),\Pi_j(\sigma ',\tau)\}=0, \\
\end{array}
     \label{esential}\ee
  it is easy to see  that
  \be \Phi^{2}_{i} \equiv \{ \Phi_i^1,H\}=\partial_{\sigma}\Pi_i|_{\sigma=0} \label{fsecondary}\ee
 \be \{ \Phi_i^1,\Phi_j^1\}=-2(MB)_{ij}\int \delta(\sigma)\delta(\sigma')\partial_{\sigma}\delta(\sigma-\sigma')d\sigma d\sigma'.
\label{pribrackets}\ee
  Similar equations also arise for $\bar{\Phi}_i$  at the
end point $\sigma=l$.  Let see how can the consistency conditions
(\ref{consistant2}) come true. Noticing (\ref{fsecondary}) and
(\ref{pribrackets}),  we find that  the two Poisson brackets
appearing on the right  hand side of equations (\ref{consistant2})
are not of the same order. This fact is explained in more detail
in  \cite{SS} using the regularized form of the Dirac delta
functions. Therefore the only way to satisfy the consistency
conditions (\ref{consistant2}) ,  is to assume that
  \be
\begin{array}{l}
  \lambda_j=\bar{\lambda}_j=0, \\
  \Phi_i^2=\bar{\Phi}_i^2=0. \\
\end{array}
   \label{rofflevel} \ee
    The important result is that the secondary constraints emerge while
the Lagrange multipliers are determined. One should continue the
consistency process by demanding the time derivatives of
$\Phi_i^2$ and $\bar{\Phi}_i^2$  to vanish. From (\ref{hamiltot})
and (\ref{rofflevel}) we have $H_T=H$ from now on, so
 \be
\Phi_i^3=\dot{\Phi}_i^2=\{\Phi_i^2,H\}=\partial^2_{\sigma}
\Phi(\sigma,\tau)|_{\sigma=0}=0
\label{consi2} \ee
  with similar expressions for $\bar{\Phi}_i^3$.
In this way two infinite constraint chains appear as
  \be
\Phi_i^n=\left \{\begin{array}{l}
  \partial^{n-1}_{\sigma}\Phi_i|_{\sigma=0}\;\;\ n=1,3,\cdots  \\
  \partial^{n-1}_{\sigma} \Pi_i|_{\sigma=0} \;\;\ n=2,4,\cdots \\
\end{array} \right.
 \label{set1}
 \ee
  \be \bar{\Phi}_i^n=\left \{\begin{array}{l}
  \partial^{n-1}_{\sigma}\Phi_i|_{\sigma=l}\;\;\ n=1,3,\cdots \\
  \partial^{n-1}_{\sigma} \Pi_i|_{\sigma=l} \;\;\ n=2,4,\cdots \\
\end{array} \right.
 \label{set2}
 \ee

Next the problem arises: what are the Dirac brackets of the fields
due to the above infinite constraints. We recall that  the Dirac
bracket of two quantities $f$ and $g$ in phase space is defined as
  \be \{f,g\}_{D.B}=\{f,g\}+\{f,\chi_i\}C^{ij}\{\chi_j,g\} \label{dirac beracket}\ee
where $\chi_i$ are second class constraints and $C^{ij}$ is the
inverse of
  \be C_{ij}=\{\chi_i,\chi_j\}. \ee
  In the present problem $C_{ij}$ is infinite dimensional and it
  is difficult (or in fact impossible) to find its inverse.
  It is obvious that for all $m$ , $n$ , $i$ and $j$
  \be \{\Phi^n_i,\bar{\Phi}^m_j\}=0. \ee
Therefore, the Dirac bracket (\ref{dirac beracket}) contains two
separate parts, one due to inverse of
$C_{ij}^{nm}=\{\Phi^n_i,\Phi^m_j\}$ and the other due to
$\bar{C}^{nm}_{ij}$, defined similarly. So it is enough to do the
calculations just for the set of  $\Phi$'s. Using the integral
form of the constraints:
  \be
\begin{array}{l}
\displaystyle{\Phi^n_i=\int d\sigma
\delta(\sigma)\partial^n_{\sigma} \Phi_i(\sigma,\tau) \hspace{1cm}
n=0,2,\cdots}\vspace{2mm}\\ \displaystyle{\Phi^n_i=\int d\sigma
\delta(\sigma)\partial^n_{\sigma} \Pi_i(\sigma,\tau) \hspace{1cm}
n=1,3,\cdots} \\
\end{array}
 \label{intform}\ee
  it is straightforward to calculate
 \be
C_{ij}^{nm}=\left\{ \begin{array}{ll}0 & n,m= 1,3,\cdots\vspace{2mm}\\
\displaystyle{-2(MB)_{ij}\int
\delta(\sigma)\delta(\sigma')\partial_{\sigma}^{m+1}\partial_{\sigma'}^n
\delta(\sigma-\sigma') d\sigma d\sigma'} & n,m= 0,2,\cdots  \vspace{2mm}\\
\displaystyle{M_{ij}\int
\delta(\sigma)\delta(\sigma')\partial_{\sigma}^{m+1}\partial_{\sigma'}^n
\delta(\sigma-\sigma') d\sigma d\sigma' }& n=1,3,\cdots\;\;\
m=0,2,\cdots
\end{array}\right.    \label{cij} \ee
This matrix possesses a finite dimensional part with $i$ , $j$
indices and an infinite dimensional  part with $m$, $n$ indices.
The finite part can be inverted easily, while for infinite part
there exist serious difficulties. In fact the most crucial point
in studying the problem is here. Among so many authors that
escaped the existence of infinite number of constraints, there are
two main references, published almost simultaneously, that have
tried to invert the matrix $C_{ij}^{nm}$ above, and calculate the
Dirac brackets. First, the famous work of Ref. \cite{CH2} in which
the authors tried to write down the inverse of $C_{ij}^{nm}$ by
using some undetermined functions $ R_{nm}(\sigma ', \sigma '')$
and $ S_{nm}(\sigma ', \sigma'')$, where it is claimed that
  these functions are omitted during calculating the Dirac brackets.
  However, we think that  it is not allowed to do calculations for the
  midpoints $\sigma'$, $\sigma''$,... of the string and thereafter
  consider them at the end points $\sigma=0, l$. In other words,
  the presence of $\delta(\sigma)$ and  $\delta(\sigma')$ in the
  formula (\ref{cij}) is essential. In fact,  it may happen that
  some expressions vanish for intermediate points of the string
before going to the boundaries. \footnote{This precise point made
the author of \cite{L} to deduce that the first level constraints
commute, and finally that the coordinate field are commutative at
the boundaries after quantization. Being more accurate
$\left[\Phi(\sigma,\tau),\Phi(\sigma',\tau)\right]=0$ for
arbitrary $\sigma$, $\sigma'$ due to antisymmetry of $B$,  while
using (\ref{intform}) shows that at boundary $\sigma=0$,
$\left[\Phi(0,\tau),\Phi(0,\tau)\right]\neq 0$ as can be seen from
(\ref{cij}).} In this way it is not clear that in  what sense may
the functions $ R_{nm}(\sigma ', \sigma'')$ and $ S_{nm}(\sigma',
\sigma'')$ be defined appropriately.

The next reference  in this regard is Ref. \cite{SS}, in which
the authors have  tried to regularize the delta functions in
(\ref{cij}) and then invert $C_{ij}^{nm}$. However, in practice
this method does not seem to make it possible to find $C^{-1}$
directly. The authors have written  the answer by considering some
desired properties of Dirac brackets. In this way the main
features of the answer is derived, but unfortunately it contains
the regularization parameter of delta functions, which does not
sound plausible. The reality is that, apart from some exceptional
references\cite{L,ZHAO},  most authors have tried to find, in
different ways,  the original results which imply the
non-commutativity of coordinate fields $X^{\mu}$ at the
boundaries (see Eqs. (\ref{endresult}) in the following). We will
show in section (\ref{section5}) that applying  the familiar
approach of mode expansion  in the context of constrained systems
gives reliable results.

Before that we prefer to spend two sections to establish the
existence of infinite number of constraints by studying the
fundamental discrete model corresponding to the continuum model.

\newsection{Discretization \label{section3}}
The discretized Lagrangian corresponding to the model given in
(\ref{original}) can be written as:
 \be L=\frac{1}{2}\epsilon \sum_{n= 0}^{N}\left(\dot{X}_n\right)^2
-\frac{1}{2}\epsilon\sum_{n=
0}^{N-1}\left(\frac{X_{n+1}-X_n}{\epsilon}\right)^{2}
+\sum_{n=0}^{N-1}\dot{X}_n B\left(X_{n+1}-X_n\right)\label{a2}
\ee
 For the sake of simplicity in notation we have dropped
the $i$, $j$ indices on $B$-field. So $B$ should be considered as
a matrix and $X_n$ as a column vector in the above equation as
well as in the following. All associated quantities to $X_n$ carry
the same hidden index of a column vector. The continuum limit is
achieved by the following replacements: \be
\begin{array}{l}
N\rightarrow \infty \\
\epsilon\rightarrow 0 \\
N\epsilon\rightarrow l \\
n\epsilon\rightarrow \sigma \\
X_n\rightarrow X(\sigma,\tau)\vspace{2mm} \\
\displaystyle{\frac{X_{n+1}-X_n}{\epsilon}\rightarrow
\partial_{\sigma}X(\sigma,\tau)}. \\
\end{array}
 \label{conlimit}\ee
 Here we have ascribed the right difference (divided by $\epsilon$)
to spatial derivative. It is also possible to do the same with the
left difference. In the continuum limit the physical quantities in
the neighboring points are the same.

The Euler-Lagrange equations of motion for intermediate points
are:
  \be \ddot{X}_n-\frac{\Delta_n}{\epsilon^{2}}+\epsilon
B\frac{\dot{\Delta}_n}{\epsilon^{2}}=0 \label{a9} \ee
  where
   \be
   \Delta_n\equiv X_{n+1}-2X_n+X_{n-1}.
   \ee
The last term in (\ref{a9}) is of order $\epsilon$ and vanishes in
the continuum limit (\ref{conlimit}) giving the wave equation
(\ref{EOM}) which implies that the $B$-field has no effect in the
intermediate points and  appears only in the equations of motion
of the boundary points, as follows
 \be
\begin{array}{l} \displaystyle{\epsilon \ddot{X}_0-
\frac{X_1-X_0}{\epsilon}-B(2\dot{X}_0-\dot{X}_1) =0 } \vspace{2mm}
\\ \displaystyle{
\epsilon\ddot{X}_N+\frac{X_{N}-X_{N-1}}{\epsilon}+ B\dot{X}_{N-1}
=0}. \label{a12} \end{array}
 \ee
In the continuum limit the terms proportional to $\epsilon$ in
Eqs. (\ref{a12}) vanish, while $\dot{X}_1$ and $\dot{X}_{N-1}$ can
be replaced by $\dot{X}_0$ and $\dot{X}_{N}$ respectively, to
obtain the following acceleration-free equations
  \be
\partial_{\sigma}X(\sigma,\tau)+B\dot{X}(\sigma,\tau)|_{\sigma=0,l}=0. \label{a13}
 \ee
Eqs. (\ref{a13}) are the primary Lagrangian constraints. In the
Hamiltonian formalism the  momenta conjugate to coordinates $X_n$
are
  \be \begin{array}{l}
p_n =\epsilon\dot{X}_n+B(X_{n+1}-X_n)\;\;\;\
n=0,1,\cdots,N-1 \\
p_N =\epsilon \dot{X}_n .\\
\end{array}
   \label{a15}\ee
The canonical  Hamiltonian reads:
 \be H={1 \over 2\epsilon}
\sum_{n= 0}^{N-1}\left[{(p_n-B(X_{n+1}-X_n))^2
+(X_{n+1}-X_n)^2}\right]+ {1\over2\epsilon}p^{2}_N. \label{a16}
 \ee
To achieve  the continuum limit one should  complete the list
given in (\ref{conlimit}) as follows:
  \be
\begin{array}{l}
p_n\rightarrow 0 \;\;\;\ n=0,1,\cdots,N  \vspace{2mm}\\
\displaystyle{\frac{p_n}{\epsilon} \rightarrow
\Pi(\sigma,\tau)=\dot{X}(\sigma,\tau)+B\partial_{\sigma}{X}(\sigma,\tau)
\;\;\;\ n=0,1,\cdots,N-1}\vspace{2mm}\\
\displaystyle{\frac{H}{\epsilon}\rightarrow {\cal H}
(\Pi(\sigma,\tau),X(\sigma,\tau),\partial_{\sigma}X(\sigma,\tau))}
\end{array}  \label{a19}
 \ee
where in the second line Eq.(\ref{a15}) is used and
$\Pi(\sigma,\tau)$ and ${\cal H}$ are the momentum field and
Hamiltonian density, respectively. The canonical equations of
motion for  the intermediate points are as follows:
 \be  \dot{p}_n=
\frac{1}{\epsilon}\left[B(p_n-p_{n-1})+M \Delta_n\right] \;\;\
n=1,2,\cdots,N-1 \label{a21a} \ee \be
\dot{X}_n=\frac{1}{\epsilon}\left[p_n-B(X_{n+1}-X_n)\right]
\;\;\;\ 0\leq n < N. \label{a21b}\ee For the end points we have
 \be
\begin{array}{l}
\displaystyle{\dot{p}_0= \frac{1}{\epsilon}\left[Bp_0+M(X_1-X_0)\right]}\vspace{2mm}, \\
 \displaystyle{\dot{p}_N=-\frac{1}{\epsilon}\left[Bp_{N-1}+M(X_N-X_{N-1})\right]. }\\
\end{array}
 \label{a23}\ee
The right hand side of the above equations are finite in the
continuum limit, while the left hand side vanishes (see Eqs.(
\ref{a19})). So Eqs. (\ref{a23})  can be viewed as the following
primary Hamiltonian \cons
  \be
 \Phi^{1}=M\partial_{\sigma}X(\sigma,\tau)+B\Pi(\sigma,\tau)|_{\sigma=0,l}  .
\label{a24}
 \ee
It should be noticed that
$(p_{\scriptscriptstyle{N}}-p_{\scriptscriptstyle{N-1}})$ is of
order $ \epsilon^2$, so in the continuum limit
$p_{\scriptscriptstyle{N-1}}/\epsilon$ can also represent the the
end point momentum $ \Pi(\sigma=l,\tau)$ as well as
$p_{\scriptscriptstyle{N}}/\epsilon$. For the same reason
$(X_{\scriptscriptstyle{N}}-X_{\scriptscriptstyle{N-1}})/\epsilon$
can be interpreted as
$\partial_{\sigma}X(\sigma,\tau)|_{\sigma=l}$ despite  our
previous convention of attribution of right difference  to the
spatial derivative at a given point. These constraints can also be
derived from Lagrangian constraints (\ref{a12}) upon inserting
$\dot{X}(\sigma,\tau)$ from (\ref{momentun}). The constraints at
the end points are completely similar to each other. So without
losing any point, we can concentrate only on the boundary
$\sigma=0$. The same arguments can be established for the other
boundary $\sigma=l$.

\newsection{Consistency condition for the constraints \label{section4}}
 In the previous section we showed that the
equations of motion for the end points of the string can be
treated as primary constraints. As in any constrained system one
should investigate the consistency of the constraints. In discrete
model this means that one should differentiate the equations
producing the constraints, i.e. Eqs. (\ref{a23}) with respect to
time to give \be
\frac{1}{\epsilon}\left[B\dot{p}_0+M(\dot{X}_1-\dot{X}_0)\right]=O(\epsilon).
\label{a100}\ee Using Eq. (\ref{a21b}) to insert $\dot{X}_0$ and
$\dot{X}_1$ into (\ref{a100}) gives
  \be
\frac{1}{\epsilon}B\dot{p}_0+\frac{1}{\epsilon^{2}}M(-B\Delta_1+p_1-p_0)=O(\epsilon).
\label{a101}
 \ee
  Inserting  $M\Delta_1$ from Eq (\ref{a21a}) in Eq
(\ref{a101}) results in
  \be
\frac{1}{\epsilon}B(\dot{p}_0-\dot{p}_1)
+\frac{1}{\epsilon^2}(p_1-p_0)= O(\epsilon). \label{b2}
 \ee
  The first term is of order $\epsilon$ while the second term is
  the discrete version of $\partial_{\sigma}\Pi(\sigma,\tau)|_{\sigma=0}$. Therefore in the
limit $\epsilon\rightarrow 0$ the second level constraint emerges
as \be \Phi^{2}=\partial_{\sigma}\Pi(\sigma,\tau)|_{\sigma=0}=0
\label{b3}\ee

One should proceed  the consistency condition for the new
constraint $\Phi^{2} $. Due to some technical difficulties if we
wish to do this in the discrete model  it is not a good idea to
differentiate directly the discrete version $
\frac{p_1-p_0}{\epsilon^2}$ of the constraint $\Phi^{2}$. Instead,
it is better to represent $\Phi^{2}$ with
$\frac{p_2-p_1}{\epsilon^2}$. In fact  we can transfer the
condition $\partial_{\sigma}\Pi(\sigma,\tau)|_{\sigma=0}=0$ to the
right by infinitesimal distance $\epsilon$. This is reasonable
since in the continuum limit every local condition on the fields
should be valid in a small neighborhood of a point, not just
strictly at the given point. This is in fact the ``continuity
hypothesis". In other words, it is not plausible  to go to
continuum limit just by taking the limits given in equations
(\ref{conlimit}) and (\ref{a19}). It is also needed to impose the
continuity hypothesis on the physical quantities, which implies
that the difference of discrete values of  fields in the
neighboring points can be at most of order $\epsilon$. Therefore,
differentiating $\Phi^{2}=\frac{p_2-p_1}{\epsilon^2}+O(\epsilon)$
gives
  \be \Phi^{3}=\frac{1}{\epsilon^2}(\dot{p}_2-\dot{p}_1)
 +O(\epsilon)\label{b5} .\ee
It is worth noting that differentiation with respect to  time does
not change the order of a quantity. The reason is that time
derivative of a quantity is achieved by the Poisson bracket of
that quantity with the Hamiltonian which is of order $\epsilon$;
but in computing the Poisson bracket, one differentiates with
respect to canonical momenta $p_n$, which are also of order
$\epsilon$. So the net result is of the same order. We insert
$\dot{p}_1$ and $\dot{p}_2$ from Eq. (\ref{a21a}) into Eq. (
\ref{b5}), to get
  \be \Phi^{3}=\frac{1}{\epsilon^3}\left[B(p_2-2p_1+p_0)
+M(X_3-3X_2+3X_1-X_0)\right]+O(\epsilon)  \label{b6} \ee
  Going to the continuum limit we have
 \be \Phi^{3}=\left[B\partial^2_{\sigma}\Pi(\sigma,\tau)
 +M\partial^3_{\sigma}X(\sigma,\tau)\right]|_{\sigma=0}=
\partial^2_{\sigma}\Phi^{1} \label{b9}\ee
 where we have used (\ref{a24}).

To investigate the consistency of $\Phi^{3}$, the strategy is the
same as before: we should differentiate $\Phi^{3}$ with respect to
time, but we need to transfer the terms one step to the right.
Then using the equations of motion (\ref{a21a}), (\ref{a21b}) and
the continuity hypothesis, and  going back to the continuum limit,
one gets the next constraint as
  \be \Phi^{4}=\partial^3_{\sigma}\Pi(\sigma,\tau)|_{\sigma=0}.\label{b10}\ee

It is reasonable that this procedure will produce at the
boundaries the infinite set of constraints
  \be\Phi^{1},\;\;\
\partial_{\sigma} \Pi,\;\;\ \partial^{2}_{\sigma}\Phi^{1},\;\;\
\partial^{3}_{\sigma}\Pi,\cdots \label{b11} \ee
 which is the same as (\ref{set1}) and (\ref{set2}). One may wonder about
the validity of \cons not only at the end points, but also in an
infinite number of their adjacent points. The answer is that a
real continues system consists of uncountably infinite points,
while in the discretized model one imposes the constraints on a
countable infinite number of points in the vicinity of the
boundaries. So physically speaking, nothing bad has happened. In
other words, suppose we extend the validity of the constraint
$\partial_{\sigma}X(\sigma,\tau)|_{\sigma=0}=0$ (in the case of a
simple string with free end points) to a large countable number of
the adjacent points of the boundary. Even when the number of
points goes to infinity, it will still remain in an infinitesimal
neighborhood of the boundary. In other words, it will never extend
in the continuum limit to a finite distance from the boundary.

\newsection{Reduced phase space \label{section5}}
In this section we try to find out the most suitable basis to
describe the physical (reduced) phase space. Whenever second class
constraints exist, one should first impose the constraints to
eliminate the redundant variables and reach the reduced phase
space. One should then try to find the most suitable bracket on
the reduced phase space. It is clear that the ordinary canonical
quantization procedure (i.e. converting the Poisson brackets to
commutators) is not consistent in the original phase space, since
quantum operators corresponding to constraints have non-vanishing
commutators which contradicts the necessity that they should
vanish either strongly or on the physical states. However, a
consistent quantization procedure can be done in the reduced phase
space. This is achieved by converting the induced brackets on the
reduced phase space to commutators.

Fortunately the famous Darboux theorem ensures us that a unique
and well defined bracket, which is the same as the Dirac bracket,
exists on this space \cite{fadjac}. In fact, the Poisson bracket
in the original phase space induces the Dirac bracket on the
reduced phase space \cite{dirac,HENOUX}. In other words, for any
two functions $f(q,p)$ and $g(q,p)$ one can write
 \be  \left\{f(q,p)\, , \, g(q,p)
\right\}_{D.B}=\left\{ f(q,p)|_{\Phi=0}\, ,\, g(q,p)|_{\Phi=0}
\right\}, \label{b14} \ee
 where $f(q,p)|_{\Phi=0}$ means evaluation of $f(q,p)$ on the
constraint surface described by the equations $\Phi=0$.

In the general case, the second class constraints may be some
complicated functions of the coordinates. If so, the constrained
and physical degree of freedom are mixed with each other and it is
not generally  an easy task to separate them. Sometimes it is
almost impossible to compute the Dirac brackets directly from the
definition (\ref{dirac beracket}), as is the case for our current
model (string in the background B-field). Moreover, it may happen
that the resulted quantum algebra is difficult to handle,
specially in order to find the corresponding representations.

Now consider an idealized model in which the second class
constraints are given by $q_{k+1}, \cdots q_{N}\, , \, p_{k+1},
\cdots p_{N}$ where the coordinates $(q_{1}\cdots q_{N}\, , \,
p_{1}\cdots p_{N})$ describe the original phase space in which the
Poisson brackets are defined as
 \be \{f,g\}=\sum_{i=1}^{N}\left( \frac{\partial f}{\partial q_i}
 \frac{\partial g}{\partial p_i}-\frac{\partial f}{\partial p_i}
 \frac{\partial g}{\partial q_i}\right) \label{Poisson}. \ee
 It is clear that the reduced phase space with
coordinates $(q_{1}\cdots q_{k}\, , \, p_{1}\cdots p_{k})$
acquires a natural bracket in which summations run from $1$ to
$k$, i.e.
 \be \{f,g\}_{D.B.}=\sum_{i=1}^{k}\left( \frac{\partial f}{\partial q_i}
 \frac{\partial g}{\partial p_i}-\frac{\partial f}{\partial p_i}
 \frac{\partial g}{\partial q_i} \right) \label{trunPoisson}. \ee
Such a {\it truncated Poisson bracket} is a realization of the
instruction (\ref{b14}) and can be checked that is equal to the
Dirac bracket of (\ref{dirac beracket}). In fact, after imposing
the constraints on this system, the Dirac brackets of the
remaining variables are the same as their Poisson brackets. Hence,
quantization of the system may be achieved in the most simple way,
such that the familiar algebra of $x-p$ variables and the
corresponding representations are still valid.

Due to extreme simplicity of the above system, it is much more
convenient to change the coordinates of a theory with second class
constraints to a basis in which the constraints constitute a set
of conjugate pairs. We call such a coordinate system as {\it
normal coordinates}. In most physical theories the Fourier modes
are normal coordinates. Traditionally people are used to impose
{\it assumed} commutation relations among the Fourier coefficients
in order to quantize a field. However, the important point, which
is not clearly stated in the literature, is that the
\textit{Fourier modes are the normal coordinates describing the
reduced phase space.} In other words, they carry the "net physics
of the theory", hence they are independent variables which are
suitable for quantization.

\subsection{Free bosonic string}For example  consider  a  free bosonic string with
Neumann boundary conditions \cite{SS}. In  this simple case  we
are given the conjugate fields $X(\sigma,\tau)$ and $
\Pi(\sigma,\tau)$ with the following Poisson bracket
 \be \{X(\sigma,\tau),\Pi(\sigma ',\tau)\}=\delta(\sigma-\sigma '). \label{c1}\ee
 Suppose the fields are real. The most general form of their Fourier expansion
can be written as: \be
\begin{array}{l} X(\sigma,\tau)=\displaystyle{\frac{1}{\sqrt{2\pi}}
\int_{-\infty}^{\infty} d k
\left[a_k(\tau)\cos k\sigma +b_k(\tau)\sin k\sigma\right]} \vspace{3mm}\\
 \Pi(\sigma,\tau)=\displaystyle{\frac{1}{\sqrt{2\pi}}\int_{-\infty}^{\infty}
  d k \left[c_k(\tau)\cos k\sigma +d_k(\tau)\sin k\sigma\right]}.
\end{array}
\label{c2} \ee

  It is obvious that $a_k(\tau)$ and $c_k(\tau)$ should
be even functions of $k$ while $b_k(\tau)$ and $d_k(\tau)$ should
be odd. Using (\ref{c1}) it is easy to see that $(a_k,c_k)$ and
$(b_k,d_k)$ are conjugate pairs in the space of new variables,
i.e. \be
\{a_k(\tau),c_{k'}(\tau)\}=\{b_k(\tau),d_{k'}(\tau)\}=\delta(k-k')
\label{c4}\ee
 and all other Poisson brackets vanish. The second class
constraints of the system are:
 \be \begin{array}{l} \partial_{\sigma}^{2n+1}X(\sigma,\tau)|_{\sigma=0, l}=0
   \vspace{2mm} \\
\partial_{\sigma}^{2n+1}\Pi(\sigma,\tau)|_{\sigma=0, l}=0
\end{array} \hspace{1.5cm} n=0,1,2,\cdots
 \label{c5}\ee
which can be derived similar to section (\ref{section2}) or by
imposing $B=0$ on its results.
 The constraints at $\sigma=0$,  give
\be
\begin{array}{l}
  \displaystyle{\int_{-\infty}^{\infty} d k(-1)^n
  k^{2n+1} b_k =0}\vspace{2mm}   \\ \displaystyle{
 \int_{-\infty}^{\infty} d k (-1)^n k^{2n+1} d_k  =0}
\end{array}
\label{c6}\ee

  Since $b_k$ and $d_k$ are odd,
 Eqs. (\ref{c6}) can be satisfied  for all $n$, iff $b_k=d_k=0$,
hence Eqs. (\ref{c2}) change to
 \be
 \begin{array}{l}
X(\sigma,\tau)=\displaystyle{\frac{1}{\sqrt{2\pi}}\int_{-\infty}^{\infty}
d k a_k(\tau)\cos k\sigma } \vspace{2mm} \\
\Pi(\sigma,\tau)=\displaystyle{\frac{1}{\sqrt{2\pi}}\int_{-\infty}^{\infty}
d k c_k(\tau)\cos k\sigma} .
\end{array}
\label{c7}
 \ee
This means that in the  basis of Fourier modes, $b_k$ and $d_k$
are constrained variables and the reduced phase space is simply
achieved by omitting them.  This is the advantage of using the
Fourier modes as normal coordinates. If we were insisting  on
working  in the original basis $X(\sigma,\tau)$ and $
\Pi(\sigma,\tau)$, we would encounter difficulties explained at
the end of section (\ref{section2}). Now let us consider the
constraints on the end point $\sigma=l$. They give \be
\begin{array}{l}\displaystyle{  \int_{-\infty}^{\infty}d k
(-1)^n k^{2n+1} a_k \sin k l=0} \vspace{2mm}  \\
 \displaystyle{ \int_{-\infty}^{\infty}d k(-1)^n k^{2n+1} c_k \sin k l=0}
\end{array}\label{c9}
 \ee
 Since $a_k$ and $c_k$ are even, the integrands in Eqs. (\ref{c9}) are
 even with respect to $k$. So the only way to impose the
 constraints is:
 \be a_k=c_k=0 \;\;\ \textrm{for}\;\;\ kl \neq0, \pi, 2\pi,...\label{c99}\ee
 Once again we see the miracle of working with Fourier modes. In
this basis a large class of the coordinates $a_k$ and $c_k$ are
omitted due to the constraints, just remaining with those with
discrete values for $k$ as $k=n\pi/l$ for integer $n$. Finally the
original field variables can be expanded in terms of infinite
countable Fourier modes $a_n$ and $c_n$ as canonical coordinates
of the reduced phase space as:
 \be \begin{array}{l} \displaystyle{
X(\sigma,\tau)=\frac{1}{\sqrt{l}}\; a_0(\tau)+\sqrt{\frac{2}{l}}\;
\sum^\infty_{n=1}a_n(\tau)\cos \frac{n\pi \sigma}{l}} \vspace{3mm} \\
\displaystyle{ \Pi(\sigma,\tau)=\frac{1}{\sqrt{l}}\;c_0(\tau)
+\sqrt{\frac{2}{l}}\;\sum^\infty_{n=1}c_n(\tau)\cos \frac{n\pi
\sigma}{l}} \end{array}\label{c11}
 \ee
  It is easy to see that the Fourier modes $a_m$ and
$c_n$ obey the canonical brackets:
  \be \{a_m,a_n\}=\{c_m,c_n\}=0,\;\;\ \{a_m,c_n\}=\delta_{mn} \label{c12}\ee
  Using expansion  (\ref{c11}) and brackets (\ref{c12}), one can compute
the Dirac brackets of any two physical functions  of the original
variables $X(\sigma,\tau)$ and $\Pi(\sigma,\tau)$. This means that
we have followed the prescription given in Eq. (\ref{c2})  to find
the Dirac brackets. As is well known \cite{dirac}, the second
class constraints should strongly vanish before quantization. This
fact can be stated in terms of the Fourier modes more clearly. The
constrained modes $(b_k , d_k)$ for all $k$ and $(a_k , c_k)$ for
$k\neq n\pi/l$ should vanish before quantization. Then one can
quantize the theory by assuming canonical commutation relations
among $(\hat{a}_m,\hat{c}_m)$ in the expansion (\ref{c11}) as
 \be
 \left[\hat{a}_m,\hat{c}_n\right]=i\hbar \delta_{mn}. \label{c13}
  \ee

\subsection{Open string in background $B$-field}
Let us now consider the string in the background
$B$-field. We expand the main fields $X(\sigma,\tau)$ and
$\Pi(\sigma,\tau)$ as done in (\ref{c2})  with $(a_k,c_k)$ and
$(b_k,d_k)$ as conjugate pairs. The constraints (\ref{set1}) and
(\ref{set2}) can be rewritten as
 \be
\begin{array}{l}\partial^{2n}_{\sigma}\Phi(\sigma,\tau)|_{\sigma=0, l}=0
\\ \partial^{2n+1}_{\sigma}\Pi(\sigma,\tau)|_{\sigma=0,
l}=0  \end{array} \hspace{1cm} n=0,1,2,\cdots\label{c15}
 \ee
 Using  the constraints (\ref{c15}) at $\sigma=0$ read
 \be
\begin{array}{l}\displaystyle{\int_{-\infty}^{\infty}d k(-1)^n k^{2n}
(k M b_k+B c_k)=0 } \vspace{2mm} \\
\displaystyle{\int_{-\infty}^{\infty}d k(-1)^n k^{2n+1} (d_k)=0 }.
\end{array}
   \label{c17}
 \ee
Remembering that $(b_k , d_k)$ are odd and $c_k$ is even with
respect to $k$, the constraints (\ref{c17}) are satisfied for all
$ n$ and $k\neq 0$ iff
 \be \begin{array}{l} \displaystyle{ b_k=
 -\frac{1}{k}M^{-1}B c_k }\vspace{2mm} \\
 \displaystyle{ d_k=0}. \end{array}\label{c19} \ee

Now imposing the constraints (\ref{c15}) on the end point
$\sigma=l$ gives
 \be \begin{array}{l}\displaystyle{\int_{-\infty}^{\infty}
d k(-1)^n k^{2n} (-k M a_k)\sin k l =0}\vspace{2mm} \\
\displaystyle{\int_{-\infty}^{\infty}d k(-1)^n k^{2n+1} ( c_k)
\sin k l =0 }.
 \end{array} \label{c21}
  \ee
These equations show that $(a_k , c_k)$ and consequently $b_k$
should vanish for $ k \neq n\pi/l$ ($n$ integer). To this end,
some care is needed for the zero mode ($k=0$). In the limit
$k\rightarrow 0$, using (\ref{c19}) we have
 \be \lim_{k\rightarrow 0} b_k\sin k\sigma = \lim_{k\rightarrow 0}
 \left( -\frac{1}{k}M^{-1}Bc_k \sin k\sigma \right)=-M^{-1}Bc_0
 \sigma . \label{r1}  \ee
Therefore the linear term $(-M^{-1}Bc_{0}\sigma )$, coming from
the zero mode of the sine terms should be present in the expansion
of $X(\sigma ,\tau )$. Similarly to (\ref{c11}) the term
$\frac{1}{\sqrt{l}}a_0$ is also necessary as the zero mode of
cosine terms. However according to the global symmetry given in
(\ref{global}), we are allowed to add any constant term to the
expansion of $X(\sigma,\tau)$. This  term should  not disturb the
validity of constraints (\ref{c15}) and should vanish in the limit
$B\rightarrow 0$. We fix this arbitrariness by adding the constant
term as $M^{-1}Bc_0 l/2$. As we will see later, $c_0$ is constant
according to the equations of motion and moreover, this choice
guarantees that the coordinates of the center of mass of the
string are commutative.

Putting all these results together, the most general form of the
fields satisfying the constraints can be written as
 \be \begin{array}{l}
 \displaystyle{X(\sigma,\tau)=\frac{1}{\sqrt{l}}\;
 \left( a_0-M^{-1}Bc_0(\sigma-\frac{l}{2}) \right) +\sqrt{\frac{2}{l}}\;
 \sum^\infty_{n=1} \left(a_n\cos \frac{n\pi \sigma}{l}
 -\frac{l}{n\pi} M^{-1}B c_n \sin \frac{n\pi \sigma}{l}\right)
  },\vspace{2mm}\\ \displaystyle{\Pi(\sigma,\tau)=\frac{1}{\sqrt{l}}
  \;c_0+\sqrt{\frac{2}{l}}\;\sum^\infty_{n=1}
 c_n\cos \frac{n\pi \sigma}{l}}. \end{array} \label{c24}
 \ee
  These relations show that in the case of mixed boundary
conditions again $a_n$ and $c_n$ are suitable canonical
coordinates of the reduced phase space. Note that $(a_n , c_n)$,
as canonical coordinates, still obey the canonical brackets
(\ref{c12}). Hence, from the general prescription given in
(\ref{b14}) it is easy to calculate the Dirac brackets of the
fields, just by using their expressions in terms of normal
coordinates $a_n$ and $c_n$. As can be seen from (\ref{c24}) the
$B$-field appears only in the expansion of coordinate fields
$X_i(\sigma,\tau)$, while the momentum fields $\Pi_i(\sigma,\tau)$
are unchanged. Therefore, the Dirac brackets
$\{X_i(\sigma,\tau),\Pi_j(\sigma ',\tau)\}_{D.B} $ and
$\{\Pi_i(\sigma,\tau),\Pi_j(\sigma ',\tau)\}_{D.B} $ are the same
as the corresponding Poisson brackets given in Eqs.
(\ref{esential}). However, for the Dirac brackets of coordinate
fields, from (\ref{c12}) and (\ref{c24}), one finds
 \be \{ X_i(\sigma,\tau),X_j(\sigma',\tau)\}_{D.B}=(M^{-1}B)_{ij}
 \left[ \frac{\sigma+\sigma'}{l}-1+\frac{2}{\pi}
 \sum_{n=1}^{\infty}\frac{1}{n}\sin\frac{n\pi}{l}
 (\sigma+\sigma')\right].
 \label{noncom}\ee
The summation over the sines is the Fourier expansion of saw waves
as follows
 \be     \sum_{n=1}^{\infty} \frac{1}{n}\sin n\theta =\left\{ \begin{array}{ll}
 -\frac{1}{2}(\pi+\theta) & -\pi\leq \theta <0 \\
 \frac{1}{2}(\pi-\theta)  & 0< \theta \leq \pi
 \end{array}\right. \ee
 This function is discontinuous at $\theta =0, 2\pi,... $.
 Supposing  its values at these points to be the average of right
 and left limits, i. e. zero, we can write the final result as:
 \be \begin{array}{l}
 \{X_i(\sigma,\tau),X_j(\sigma',\tau)\}_{D.B}=0  \hspace{2cm}
 \sigma ,\sigma'\neq 0 \\
 \{X_i(0,\tau),X_j(0 ,\tau)\}_{D.B}=-2(M^{-1}B)_{ij}, \\
 \{X_i(l,\tau),X_j(l ,\tau)\}_{D.B}=2(M^{-1}B)_{ij}. \\
 \end{array} \label{endresult}\ee
As we see, after quantization the coordinate fields are
noncommutative at the end-points of the string, in agreement with
the well-known results given in the literature \cite{AAS3, CH2,
SS}. If we had not added the constant term $M^{-1}Bc_0 l/2$ to the
expansion (\ref{c24}) the above result would have differed from
(\ref{endresult}) just by a constant term throughout the string,
as well as at the end-points. In other words the non-commutativity
at the end-points $\sigma =0$ and $\sigma =l$ have opposite signs
since we have imposed the condition  that the center of mass
coordinates are commutative.

Our emphasize in our derivation of the important result
(\ref{endresult}) is that we have not used the expansions of
fields in terms of the solutions of the equations of motion. In
fact, we have not considered the time dependence of the physical
modes $a_n(\tau)$ and $c_n(\tau)$ which should be determined by
means of the special form of the Lagrangian or Hamiltonian. This
feature of our approach will be explained more in the next
section.

\newsection{Quantization and equations of motion\label{section6}}
The traditional canonical quantization procedure  is as follows:
one considers the general solution of the classical equations of
motion, then \textit{imposes} the boundary conditions to decrease
the number of possible modes \footnote{The classical equations of
motion are usually linear differential equations; so one can
expand their solutions using a complete set of modes}, and finally
\textit{assumes} suitable commutation relations amongst the
physical modes to quantize the theory. One may ask: ``is it really
necessary, or even allowed, to use classical equations of motion
in the process of quantization?".

Let us first study the problem in an ordinary (unconstrained)
system. The important point is that the special form of the
Hamiltonian (or Lagrangian) is not the essential point that
determines the algebra of physical observables and consequently
the structure of the space of physical states of the theory. On
the other hand, the role of the Hamiltonian is just determining
the dynamics of the system. Given the initial state of a system,
the Hamiltonian is the main tool which gives the time evolution of
the state of the system. However, people usually try to construct
the basis of the space of the physical states in such a way that
the Hamiltonian operator is diagonal; since this provides
consequent convenience to follow the dynamics of the system.

For example in one dimension whatever the Hamiltonian is, one can
use the algebra of the $x-p$ operators acting on the space with
$|x'>$'s or $|p'>$'s as the basis, as well as the equivalent
algebra of $a-a^\dag$ operators and the corresponding basis in
which the operator $a^\dag a$ is diagonal. However, it is
well-known that for a free particle the former algebra is more
suitable while for the harmonic oscillator the latter is more
appropriate to study the dynamics of the system.

To this end, we want to emphasize that conceptually it is not
needed to treat the quantum fields as the solutions of the
classical equations of motion. However, the reader may have
encountered several books or papers where the authors write down
the fields as an expansion in terms of solutions of the classical
equations of motion (for example plane  waves with definite
$\omega-k$ relations) and then quantize the theory by imposing
assumed canonical algebra among the coefficients of the expansion.
A careful notice leads to the observation that most of the time,
the explicit time dependence of the terms in the expansion are not
used during the subsequent analysis. For example in harmonic
oscillator problem, it is just the quantum algebra $[a,a^\dag]=1$,
coming from the classical algebra $\{x,p\}_{P.B.}=1$ upon
quantization, which determines the basis of physical states as
$|n>,|n+1>,\cdots$.\footnote{Note that imposing the condition of
unitarity on the physical states, restricts $n$ to positive
integer values, which in this case guarantees that the energy
states are bounded from bellow. Therefore, besides the quantum
algebra of observables, some other physical requirements such as
unitarity principle play important roles in determining the set of
physical states.} Then the explicit form of the Hamiltonian may be
used to determine the time dependence of $a(t)$ and $a^\dag (t)$
in Heisenberg picture as $a(t)=a(0)\exp (-i\omega t)$ and
$a^\dag(t) =a^\dag (0)\exp (i \omega t)$, which is not essential
in determining the quantum properties of the observables as well
as the space of physical states.

Our experience shows that in quantum field theory the Fourier
expansion of the fields come out, most of the time, to be useful
in the process of quantization. As discussed in the case of the
models considered in this paper, this is just a change of
variables in the phase space from $X(\sigma,\tau)$ and
$\Pi(\sigma,\tau)$ say, to $a_k(\tau),b_k(\tau),\cdots$ etc. Then,
regardless of the dynamics of the system, the Poisson brackets of
the original variables determine those of the new ones. There are
two main advantages in this change of variables. First, the
Hamiltonian may be diagonal or have a simpler form in the new
basis. Second, the constraints as well as boundary conditions
(which are also considered as constraints in our approach) may be
applied in a simpler way in the framework of the new variables.
Therefore, the Hamiltonian has some partial role in quantization
since the dynamics of constraints, (i.e. the consistency of the
constraints) should be investigated classically \textit{before}
quantization. It is not possible to construct a quantum algebra
among the variables, unless the Hamiltonian has vanishing brackets
with the constraints on the physical space (reduced phase space).
\footnote{If  the constrained quantities are assumed as vanishing
operators in the quantization procedure, then their brackets with
the Hamiltonian should also vanish. On the other hand, if one
quantizes the theory by imposing the condition that quantum
operators corresponding to constraints should kill the physical
states, then again it is clear that their brackets with the
Hamiltonian should also kill the physical states.}

In other words, although the full dynamics of the physical
variables is not essential for quantizing the theory, the dynamics
of the constrained variables should be worked out completely
before quantization, so that the final brackets of the Hamiltonian
with the constraints vanish. This means that using the classical
equations of motion, all the secondary constraints should be
computed before quantization. Schematically  we can say
 \be
\textrm{primary constraints} + \textrm{classical equations of
motion}\longrightarrow \textrm{secondary constraints.}
\label{textformula}\ee
 This point can be seen  clearly in the example of the bosonic string
with Neumann boundary conditions. If one had considered just the
primary constraints $\partial_{\sigma}X(\sigma,\tau)|_{\sigma=0,
l}=0$ instead of
 the whole set of (\ref{c5}), then one would not have been able to deduce the
 expansion  (\ref{c11}) for the fields. However,
 if one considers the primary constraints together with the
equations of motion (resulting from the Hamiltonian
(\ref{hamiltoni}) with $B=0$) one obtains
 \be \begin{array}{l}\displaystyle{
   \partial_{\tau}X(\sigma,\tau)=\Pi(\sigma,\tau)} \\
  \displaystyle{\partial_{\tau}\Pi(\sigma,\tau)=\partial_{\sigma}^2X(\sigma,\tau) }. \\
 \end{array}
   \label{c101}\ee
 Then one can easily check that
 \be \begin{array}{l}
    \partial_{\tau}(\partial_{\sigma}X(\sigma,\tau))=\partial_{\sigma}(\partial_{\tau}X(\sigma,\tau))=
  \partial_{\sigma}\Pi(\sigma,\tau) \\
   \partial_{\tau}(\partial_{\sigma}
\Pi(\sigma,\tau))=\partial_{\sigma}(\partial_{\tau}
\Pi(\sigma,\tau))=\partial_{\sigma}^{3}X(\sigma,\tau) . \\
  \end{array}   \label{canoniceof}\ee
  In this way the infinite set of
 constraints(\ref{c5}) are in fact resulting from the
 primary constraints plus the equations of motion (see
 (\ref{textformula})). This argument shows that there is no way to
 escape the fact that an infinite number of constraints really exist.
 If one uses the full capacity of the classical equations of
 motion (\ref{c101}), or the explicit form of the Hamiltonian in
 terms of the normal coordinates as
 \be
H=\frac{1}{2}\sum_{n=0}\widetilde{c}_n
c_n+\frac{1}{2}\sum_{n=1}\widetilde{a}_n a_n(\frac{n\pi}{l})^2,
\label{hamiltonmode1}\ee
  in order to determine the dynamics
 of the physical  variables $a_n(\tau)$ and $c_n(\tau)$, then one
 obtains
$\dot{a}_n=c_{n}$ and $\dot{c}_n=-(n\pi/l)^{2}a_{n}$, which
acquire the solution \be
 \begin{array}{l}
  \displaystyle{ a_n(\tau)=a_n(0)\cos \left( \frac{n\pi }{l}\tau \right)+\frac{l}{n\pi}\,
  c_n(0)\sin \left( \frac{n\pi }{l}\tau \right) }\vspace{3mm}\\
  \displaystyle{  c_n(\tau)=c_n(0)\cos \left( \frac{n\pi }{l}\tau \right)-
  \frac{n\pi}{l}\,a_n(0)\sin \left( \frac{n\pi }{l}\tau \right ) }. \\
 \end{array}   \label{c103}\ee
However, we insist again that the full dynamics of the physical
variables is not necessary to quantize the theory. It seems that
this partial role of the classical equations of motion in
determining the dynamics of the constraints (or boundary
conditions in most of the familiar physical systems) is the hidden
reason behind the common practice of expanding the fields in terms
of the classical solutions of equations of motion before
quantization.

Now let us consider again the string in background $B$-field to
observe the above points. In this case the equations of motion
resulting from the Hamiltonian(\ref{hamiltoni}) read \be
\begin{array}{l}
   \partial_{\tau}X(\sigma,\tau)=\Pi(\sigma,\tau)-B\partial_{\sigma}X(\sigma,\tau) \vspace{2mm}\\
   \partial_{\tau}\Pi(\sigma,\tau)=B\partial_{\sigma}\Pi(\sigma,\tau)+M\partial^2_{\sigma}X(\sigma,\tau). \\
 \end{array} \label{beom}\ee
 This gives \be \begin{array}{c}
  \partial_{\tau}\left(\Phi(\sigma,\tau)\right)=\partial_{\sigma}(\Pi(\sigma,\tau)) \\
  \partial_{\tau}(\partial_{\sigma}\Pi(\sigma,\tau))=\partial_{\sigma}^2(\Phi(\sigma,\tau)) \\
  \vdots
\end{array}
 \ee In this way the full set of infinite constraints
(\ref{c15}) emerge as the result of combining  the primary
constraints $\Phi(\sigma,\tau)|_{\sigma=0,l}=0$ with the equations
of motion (\ref{beom}). Using the expansion (\ref{c24}), the
canonical Hamiltonian (\ref{hamiltoni}) in terms of the normal
coordinates can be calculated as \be
H=\frac{1}{2}\sum_{n=0}^{\infty}\widetilde{c}_n M^{-1}
c_n+\frac{1}{2}\sum_{n=1}^{\infty}\widetilde{a}_nM
a_n(\frac{n\pi}{l})^2. \label{hamiltonmode}\ee
 We observe again that the full content of
the dynamics is not needed for the quantization process. In fact,
using the Hamiltonian (\ref{hamiltonmode}), or the equations of
motion (\ref{beom}), to determine the dynamics of the physical
variables $a_n(\tau)$ and $c_n(\tau)$,  one finds
 \be \begin{array}{l}\dot{a}_n(\tau)=M^{-1}c_n(\tau)\vspace{2mm}\\
 \displaystyle{\dot{c}_n(\tau)=-\left( \frac{n\pi}{l} \right)^2M
 a_n(\tau) }\\ \end{array} \hspace{1cm} n=0,1,\cdots \label{c1111} \ee
  which acquire the solution
 \be
 \begin{array}{l}
  \displaystyle{ a_n(\tau)=a_n(0)\cos \left( \frac{n\pi }{l}\tau \right)
  +\frac{l}{n\pi}\,
  M^{-1}\,c_n(0)\sin \left( \frac{n\pi }{l}\tau \right) }\vspace{3mm}\\
  \displaystyle{  c_n(\tau)=c_n(0)\cos \left( \frac{n\pi }{l}\tau \right)-
  \frac{n\pi}{l}\,M\,a_n(0)\sin \left( \frac{n\pi }{l}\tau \right ) } \\
 \end{array}   \label{c110}\ee
 and
 \be
 \begin{array}{l}a_0(\tau)=M^{-1}c_0(0)\tau+a_0(0)\vspace{3mm} \\
  c_0(\tau)=c_0(0). \\ \end{array}   \label{c111}
 \ee
 As is apparent in order to quantize the theory, specially finding
 the important results of (\ref{endresult}), one does not need to know
 the explicit time dependence given in Eq. (\ref{c110}). Also note
 that $c_0$ is constant, in agreement with our previous trick of
 adding the constant term $M^{-1}Bc_0 l/2$ to the expansion of coordinate
 field in Eq. (\ref{c24}).

\section{Conclusion \label{section7}}
In this paper we discussed different aspects of the idea of
considering the boundary conditions as Dirac constraints.  Our
theoretical laboratory for this aim was `` an open string in a
background $B$-field". We observed that besides the singularity of
the Lagrangian, the boundary conditions can serve as a source of
introducing  the primary constraints. Analyzing in detail the
discretized version of the model shows that the primary
constraints are  the continuum limit of the equations of motion
for the end points; while the secondary constraints are derived by
imposing the consistency conditions on these equations and  then
going to the continuum limit. In this process the continuity
hypotheses  is important. This  implies that in order to get a
continuum solution, the fields in the adjacent points in the
corresponding discrete model should not differ drastically. In
this way it turns out that the discretized model highly supports
the existence of infinite chains of second class constraints.

The continuity hypotheses is also deeply related to Fourier
expansion, as follows.  It is well-known that in writing any
field as a summation over the set of well-behaving   continuous
sine  and cosine  functions, any finite discontinuity  in the
field or in a finite number of its derivatives, is not seen by the
expansion and is somehow removed from the problem. However, the
boundary conditions should not be considered as such
discontinuities. Although boundary conditions imposes some
restrictions just on definite points  at the border of the medium
(i.e. end points of the string), they have their considerable
influence throughout the whole medium. The important point is that
the Fourier expansion plays the role of a carrier of boundary
conditions from the boundaries through the medium. In fact, the
emergence of an infinite number of constraints causes serious
restrictions on the Fourier modes invited to the expansion of the
fields. Since the Fourier modes are alive in the whole medium as
continuous and well-behaved function, \textit{the message of
boundary conditions} is distributed in this way throughout the
system. The familiar example in this regard is an open string with
Neumann boundary conditions, in which the fields are expanded in
terms of a set of discrete cosine modes only. In this case the
Fourier expansion is, in fact, used to soften and flatten the
fields undergoing definite conditions at the boundaries.

Summarizing, the continuity hypotheses implies, in the discrete
model, the validity of constraints should be spread in a set of
countable infinite number of adjacent points near the boundaries.
This fact then shows itself in emerging infinite number of
constraints; and finally causes omitting a large class  of Fourier
modes, which leads somehow to propagation of the effect of
boundary conditions through the medium.

Another aim of this paper was studying the Poisson structure of
the reduced phase space.  Using the original coordinates of phase
space implies serious  difficulties in computing the complicated
Dirac brackets. We observed that Fourier modes can serve  as the
normal coordinates of the reduced phase space. Using the Fourier
modes make it possible to do  calculations. Moreover, it gives a
valuable understanding of the Poisson structure of the reduced
phase space which contains the true physical degrees of freedom
of the model. In this way after disappearing some Fourier modes
as redundant (constrained) coordinates, the remaining modes can
be viewed as the physical degrees of freedom, i.e. the
coordinates of the reduced phase space. Fortunately these modes
emerge as conjugate pairs with a simple and well-established
bracket. As explained in the text, the brackets of the remaining
modes define  the Dirac brackets of fields. In this way one
expands all associated fields and quantities just in terms of the
normal coordinates, and then using their  brackets one writes
down all the  Dirac brackets. It turns out that for an open
string with mixed boundary conditions there remains no doubt
about the fact that the  Dirac brackets of the coordinate fields
at the boundaries of the string are nonzero due to the $B$-field.
Then, upon quantization, the coordinates of the string, and hence
the coordinates of the D-brane, are non-commutative.

Another new feature in our approach is that we do not use of
solutions of equations of motion in the process of quantization.
We showed clearly, in the model under consideration, that it is in
fact possible to find the algebraic structure of the quantum
theory without any need to expand the fields in terms of solutions
of classical equations of motion. We argued that in any quantum
theory, constructed upon quantization of a classical model, one
only needs to consider the dynamics of the constraints before
quantization. In other words, it is not essential to find the
dynamics of the complete set of physical variables to do
quantization.

We think that our approach may  be useful in analyzing the
physical structure together with quantization of every model with
complicated boundary conditions. Two recent examples can be seen
in  \cite{JJ,JJ1}. This approach may be applied  also to more
complicated systems such as membranes \cite{KAWA,DAS,baner2}.

 \textbf{Acknowledgements}

The authors would like to thank M. M. Sheikh-Jabbari   for his
useful discussions and comments and A. E. Mosaffa for reading the
manuscript.

\end{document}